\documentclass[twocolumn,aps,prb,showpacs]{revtex4}
\usepackage{graphicx}
\usepackage{amsmath}

\begin{document}

%
% title page
%

\title{Assessment of a GW-BSE approximation scheme
on an asymmetric two-dimensional
interacting electron system in a perpendicular magnetic field}

\author{Xiaoguang Wu}

\affiliation{Independent Researcher, P.O.Box 912, Beijing, 100083, China}

\begin{abstract}

A GW-BSE approximation scheme is assessed by applying it to
a model of asymmetric two-dimensional (2D) interacting electron system.
The model is assumed to have a parabolic band
characterized by two independent effective mass parameters.
A perpendicular magnetic field is applied to the asymmetric 2D electron
system, and the well-known Kohn's theorem is still valid, i.e., the cyclotron
resonance is not affected by the electron-electron interaction.
This theorem imposes a constraint on the approximation scheme employed
in the treatment of electron-electron interaction.  In the present study,
the Green's function is calculated in the self-consistent Hartree-Fock
approximation.  The electron density correlation function is calculated by
solving a Bethe-Salpeter equation (BSE) in the ladder diagram approximation.
It is found that, the excitation frequency near the cyclotron resonance
frequency approaches a value that is lower than the cyclotron resonance
frequency at small wave vectors, when two effective masses are different.
When two effective masses are the same, the excitation frequency
approaches the cyclotron resonance frequency at small wave vectors as required.
Our findings suggest that the approximation scheme used in this theoretical
investigation fails to satisfy the requirement due to the Kohn's theorem,
and one should go beyond this approximation scheme.

\end{abstract}


\maketitle

%
% text
%

\section{Introduction}

The influence of electron-electron interaction on the electronic
properties of condensed matters has been studied both theoretically
and experimentally for decades.  Many novel and exotic properties
are found to arise from the intricate electron-electron
interaction \cite{martin1, martin2}.

Tackling the effect of electron-electron interaction is a challenging
and difficult theoretical task.  During the past decades, various
theoretical methods, e.g., the density functional theory, the Green's
function method or the many-body perturbation theory, and the quantum
field and quantum statistics theory, have been invented and applied
to explore various interacting systems \cite{fetter, mahan, nagaosa, nozieres}.

\begin{figure}[ht]
\includegraphics[angle=0, width=7.5truecm]{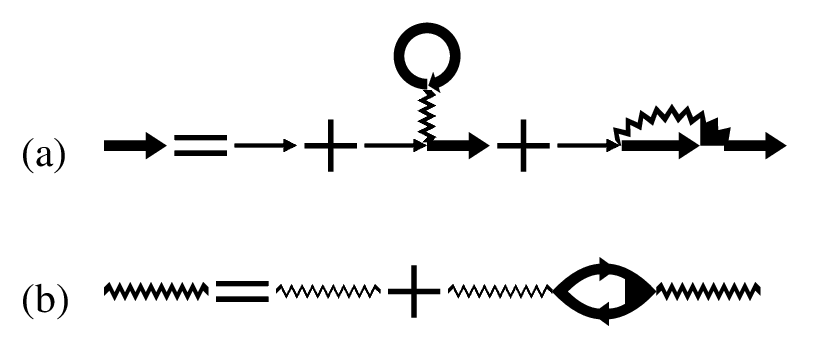}
\caption{
(a) Equation for Green's function.
(b) Equation for screened interaction.}
\end{figure}

In recent years, as the computation codes and computational resources
become more widely available, the so-called GW method has been frequently
used to study the electronic excitations in condensed
matters \cite{gw01, gw02, gw03, gw04, gw05}.
As its name implies, the GW method is a variation of Green's
function method with some specific approximations.
For an interacting electron system, the exact
Green's function and exact screened interaction satisfy two integral
equations, shown in Fig.1a and Fig.1b, respectively \cite{fetter, mahan}.
Two integral equations introduce a self-energy $\Sigma$, a screened
electron-electron interaction $W$, and a polarization function $\Pi$.
$\Sigma$, $W$, and $\Pi$ are functionals of the Green's function $G$.
In a practical GW calculation, one evaluates approximately the self-energy
to obtain the Green's function.  One also approximately calculates the
polarization function to obtain the screened electron-electron
interaction.  In those calculations, certain consistence is required.
Different approximation schemes have been proposed and employed in
practical GW implementations and their
applications \cite{gw01, gw02, gw03, gw04, gw05, gw06, gw061, gw07, gw08, gw09, gw10}.

It is not easy to assess the correctness of a particular GW approximation
scheme, as it can be viewed as a subjective selection and partial summation of some
Feynman diagrams.  In the corresponding theoretical calculations, numerical
works are heavily involved.  One usually faces computational demanding
multi-dimensional numerical integrations, and may have to sacrifice
numerical accuracy for the computational time cost.
One way to check the validity of a theory is to compare the
calculated electronic properties against experimental observations.
However, the theoretical calculation is usually based on a model system,
which does not take into account all aspects of a real material, e.g.,
the influence of disorder and electron-phonon interaction.
Issue like this makes the comparison not as strongly convincing as it should be.
Therefore, it is always desired for one to check some internal consistences
of the theory itself, when it is possible.

The preset work is motivated by the concern discussed above.
We will consider a simple model of asymmetric two-dimensional (2D) interacting
electron system.  A perpendicular magnetic field is applied to the 2D
electrons.  In this model system, the well-known Kohn's theorem is still
valid \cite{kohn}, and it provides an exact result that can be used to
examine the validity of the approximation scheme employed.

This paper is organized as follows: in section II,
the formulation is presented briefly.  In section III,
calculated results are presented and discussed.
The last section contains a brief summary.

\section{Formulations and calculations}

The Hamiltonian of asymmetric 2D interacting electron system can be
written as
\begin{align*}
H & = \sum_{\alpha} \int d{\bf r}
  \psi^{\dagger}_{\alpha}({\bf r})
  H_0 \psi_{\alpha}({\bf r})
  \\
  & + \frac{1}{2}
  \sum_{\alpha,\beta}
  \int d{\bf r} d{\bf r}'
  \psi^{\dagger}_{\alpha}({\bf r})
  \psi^{\dagger}_{\beta}({\bf r}')
  v({\bf r} - {\bf r}')
  \psi_{\beta}({\bf r}')
  \psi_{\alpha}({\bf r}) ,
\end{align*}
with $v({\bf r})=e^2/(\epsilon_0 r)$, and
\[
H_0 = \frac{1}{2m_1} ( p_x + \frac{eA_x}{c} )^2
  + \frac{1}{2m_2} ( p_y + \frac{eA_y}{c} )^2 .
\]
The 2D electrons are confined in the $xy$ plane.
${\bf A}$ is a vector potential and it generates a
uniform and perpendicular magnetic field with strength $B$.
In this paper, ${\bf A}=(0, Bx, 0)$ is used.
The model electron is assumed to have a parabolic band, but with two
effective mass parameters.  When $m_1 \ne m_2$, the system is
asymmetric.

The single-particle energy levels and wave functions can be
exactly obtained \cite{merzbacher, xgwu, note}.  The single-particle energy
level is given by $\varepsilon_{l}=(l+1/2)\hbar\omega_c$
with $\omega_c=eB/(\sqrt{m_1m_2} c)$.  The magnetic field
introduces a length scale $l_B=(\hbar c/(eB))^{1/2}$.
The strength of electron-electron interaction is
characterized by the parameter $v_e$ defined as
$v_e=\sqrt{2}(e^2/(\epsilon_0 l_B))/(\hbar\omega_c)$.
In this paper, $\hbar\omega_c$ is used as the energy scale, and $l_B$
as the length scale.

\begin{figure}[ht]
\includegraphics[angle=0, width=7.5truecm]{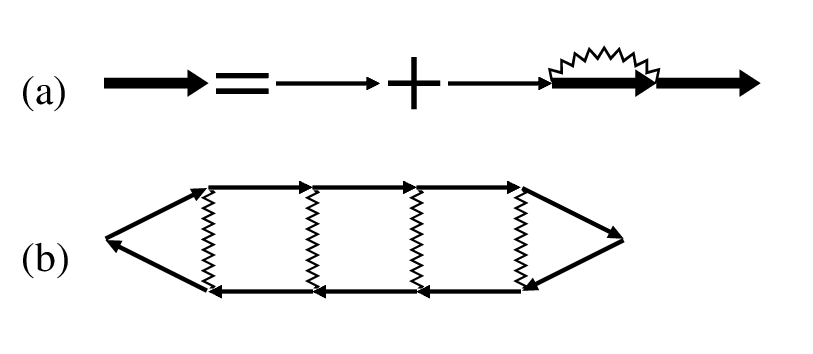}
\caption{
(a) Equation for the Green's function in the
self-consistent Hartree-Fock approximation.
(b) Diagram for the fourth order term contributed to the
irreducible density correlation function in the ladder approximation.}
\end{figure}

The one-particle Green's function is calculated within the self-consistent
Hartree-Fock approximation.  Because of asymmetry, Landau levels are mixed.
The electron density correlation function is then evaluated by
solving a Bethe-Salpeter equation in the ladder diagram approximation,
where the involved Green's functions are dressed ones.
In Fig.2a, the equation for the Green's function is shown.
In Fig.2b, the forth order contribution to the irreducible electron
density correlation function is shown.
All ladder diagrams are taken into account.

We have carried out the calculation with two form of electron-electron
interaction.  One is the bare interaction, and the other one is a statically
screened one.  The dielectric function for the static screening is taken
as the dielectric function of an isotropic 2D electrons in the presence of
a perpendicular magnetic field in the random phase approximation.
This approximation is chosen for numerical simplicity.
The overall behavior of calculated results is qualitatively the same
and we will present results calculated with the bare interaction.

\section{Results and discussions}

The calculation is done for the zero temperature case.  The occupation
of Landau levels is spin resolved, with spin-down states to be occupied first.
The excitation frequencies or excitation energies are determined by the poles
in frequency domain of the electron density correlation function.
We will focus on the branch of excitation frequency near $\omega_c$.

\begin{figure}[ht]
\includegraphics[angle=90, width=8.truecm]{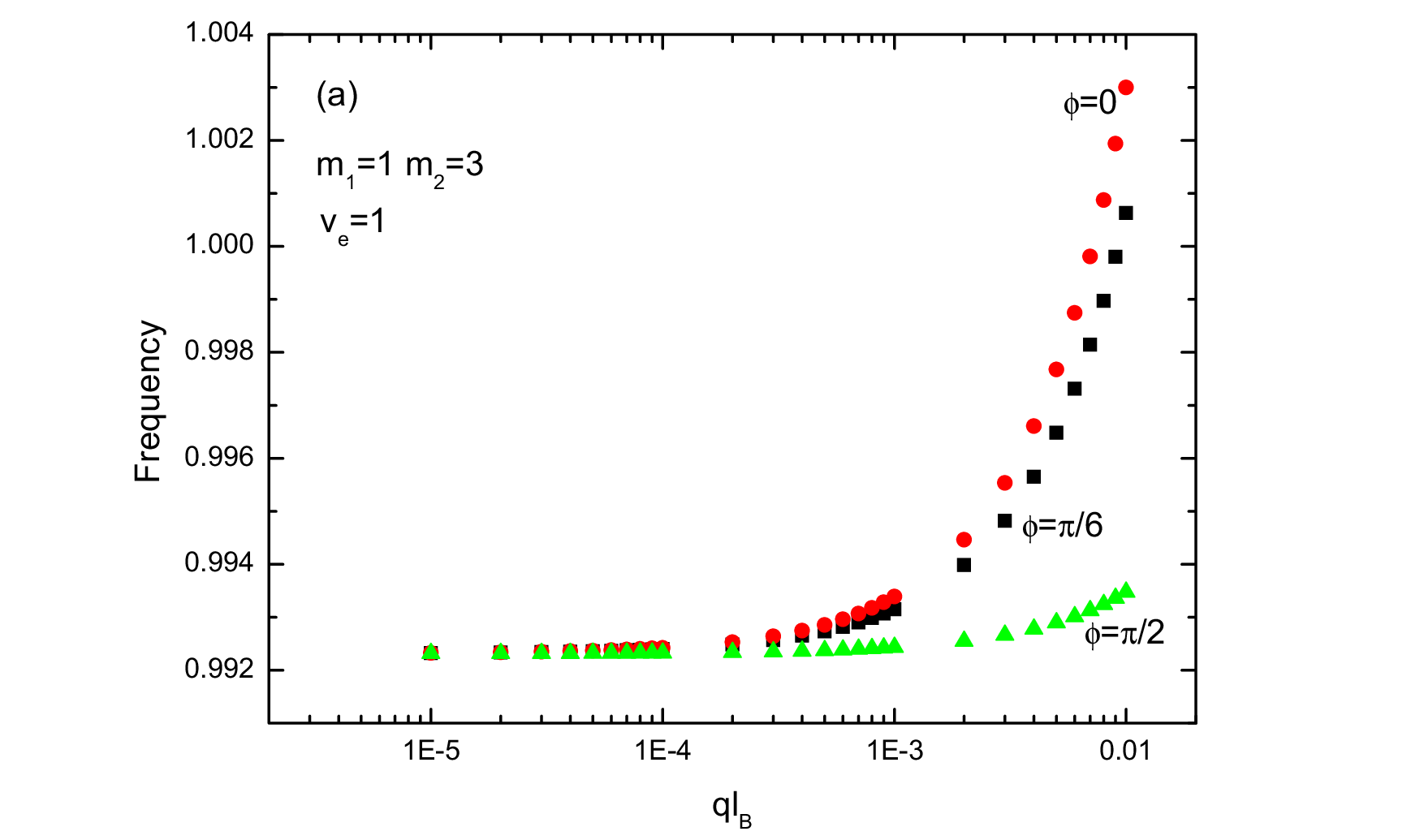}
\includegraphics[angle=90, width=8.truecm]{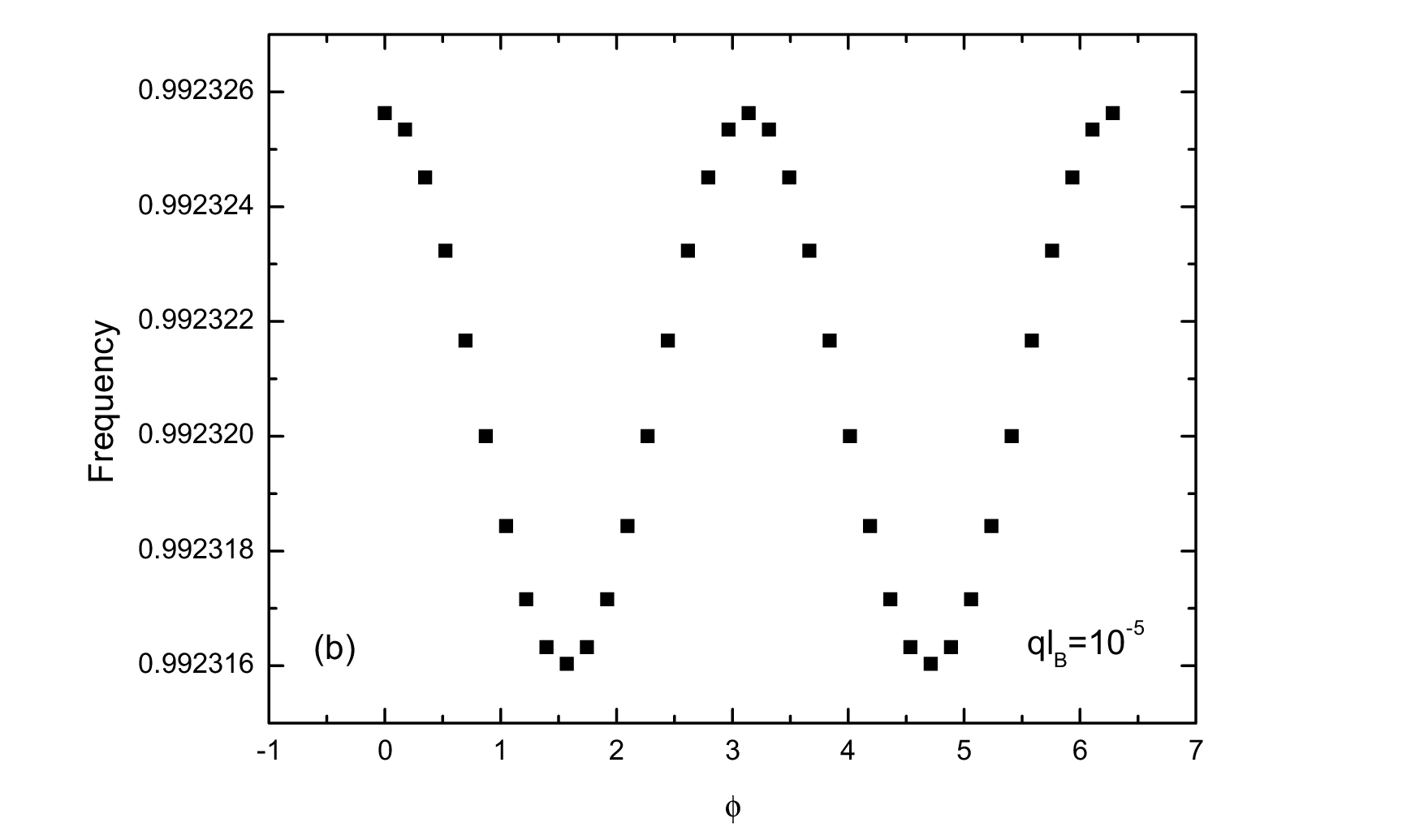}
\caption{
(a) The excitation frequency versus the magnitude of the wave vector.
(b) The excitation frequency versus the direction of the wave vector at $ql_B=10^{-5}$.
The filling factor is $1$.  $m_1=1$, $m_2=3$, and $v_e=1$. }
\end{figure}

In Fig.3a, the excitation frequency near $\omega_c$ is shown as a function of wave vector.
The excitation frequency depends on the direction of the wave vector.
An angle $\phi$ can be introduced via $q_x=q\cos\phi$.  This $\phi$ dependence
may be traced back to the wave vector dependence in the matrix element of
electron-electron interaction.
The matrix element $|\langle \lambda'| e^{i{\bf q}\cdot{\bf r}} | \lambda \rangle|$
is a function of $q[(m_1/m_2)\cos^2\phi+(m_2/m_1)\sin^2\phi]^{1/2}$.

At a larger $q$ value, the excitation frequency is obviously $\phi$ dependent,
and is larger than $\omega_c$.  As the value of $q$ decreases, one observes in Fig.3a that,
the excitation frequencies apparently approach to one value that is clearly
smaller than $\omega_c$.  Currently, no analytical equation is available for the
excitation frequency in the ${\bf q}=0$ limit.  One can not calculate the
excitation frequency at ${\bf q}=0$ directly, due to a numerical divergency.
Therefore, one has to rely on numerical calculation methods to explore the region
of very small wave vectors.

In Fig.3b, the excitation frequency versus $\phi$ is shown.  $ql_B=10^{-5}$ is fixed.
As one can see from the vertical scale of Fig.3b, this $\phi$ dependence
becomes quite weak at small wave vectors.  But the $\phi$ dependence can
still be clearly seen.  Our calculated data points can be connected and form a quite
smooth curve.  This demonstrates the high accuracy achieved in our
numerical calculation.  Because of this high numerical accuracy, one can confidently
trust the calculated excitation frequency at small wave vectors.

\begin{figure}[ht]
\includegraphics[angle=90, width=8.truecm]{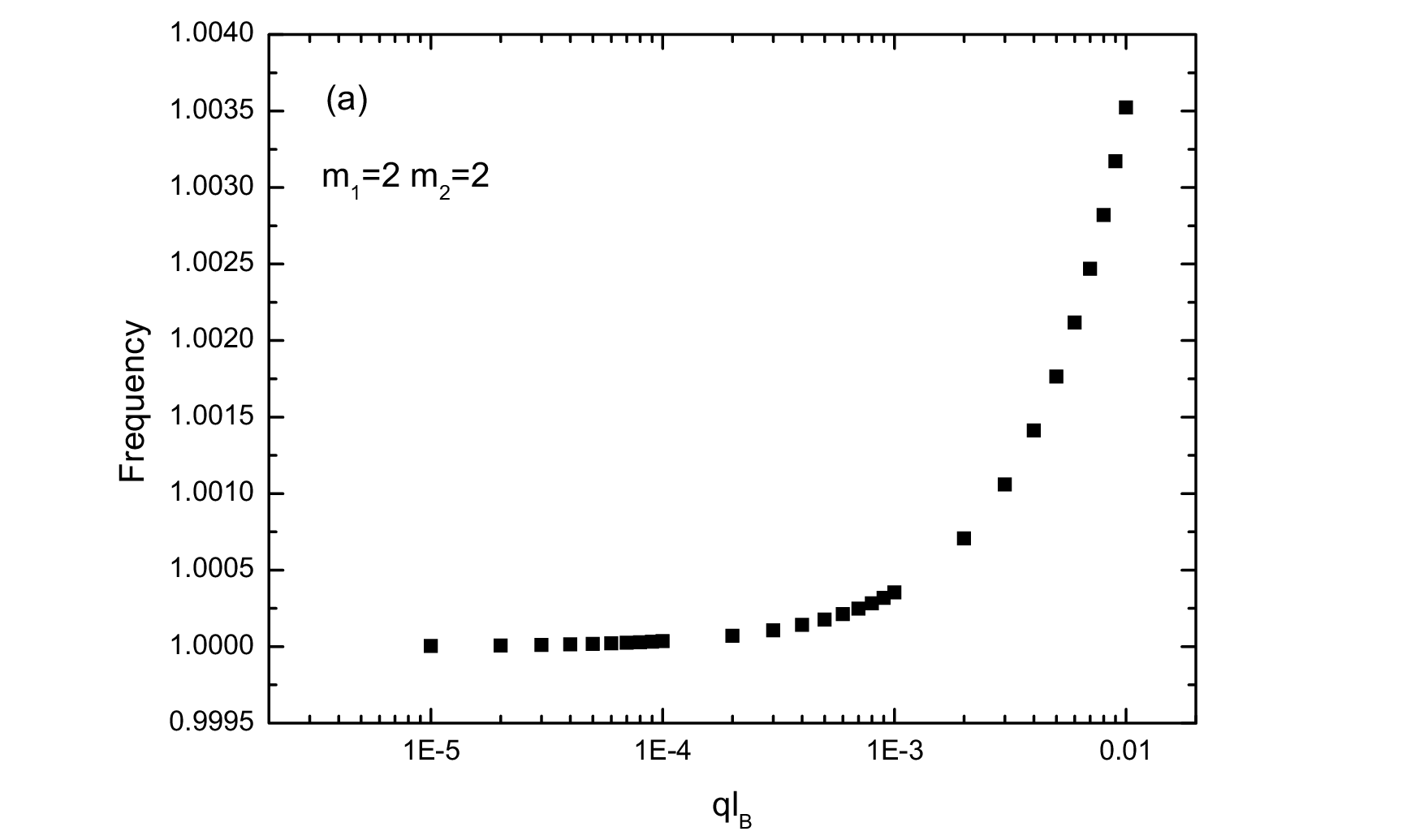}
\includegraphics[angle=90, width=8.truecm]{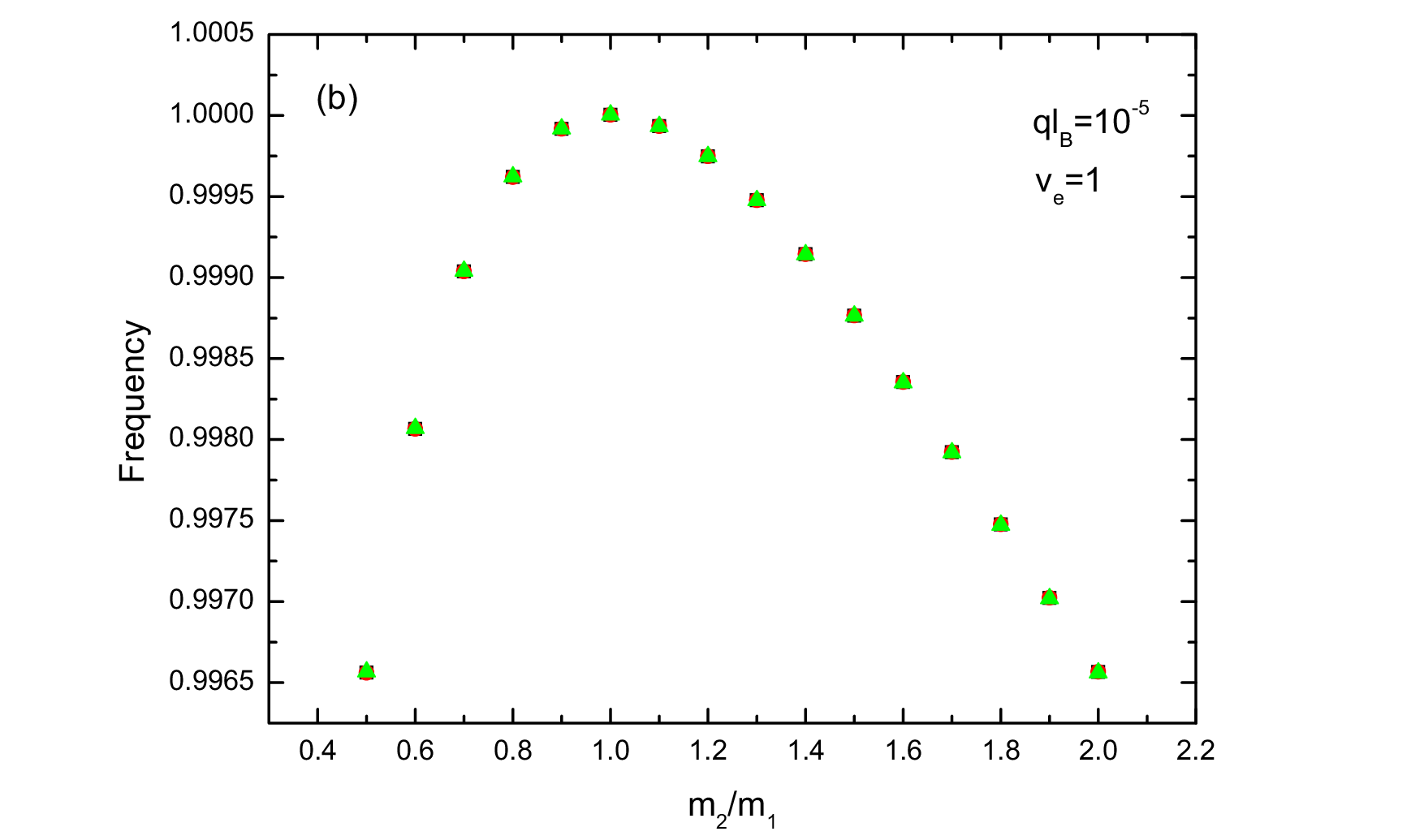}
\caption{
(a) The exicitation frequency versus the magnitude of the wave vector for $m_1=m_2=2$.
(b) The excitattion frequency versus $m_2/m_1$ at $ql_B=10^{-5}$.
The filling factor is $1$, and $v_e=1$. }
\end{figure}

In our calculation, only the ratio of $m_1/m_2$ matters, but we show the
individual value of $m_1$ and $m_2$ in the figure.  There should exist no
confusion.  The filling factor is assumed to be $1$, in Fig.3a and Fig.3b.
Therefore, only one Landau level is occupied.  In this case, the excitation
frequency near $\omega_c$ has only one branch.  When the filling factor
is $2$, there will be two branches of excitation frequency near $\omega_c$.
The strength of electron-electron interaction is assumed as $v_e=1$.

In Fig.4a, the excitation frequency versus the wave vector is shown,
however, with two effective mass parameters having the same value.
In this case, the excitation frequency has no angle $\phi$ dependence.
One also observes that, as the value of $q$ decreases, the excitation frequency
approaches $\omega_c$ from above.  This behavior is expected.  The same behavior
had been obtained in many previous theoretical calculations.

In Fig.4b, the excitation frequency is shown as a function of $m_2/m_1$,
at fixed $ql_B=10^{-5}$.  The filling factor takes a value of $1$.
The electron-electron interaction strength $v_e=1$ is also assumed.
One observes that, the excitation frequency approaches $\omega_c$
when $m_2/m_1$ approaches $1$.  When $m_2/m_1$ deviates away from $1$,
the excitation frequency becomes lower than $\omega_c$.
In Fig.4b, the calculation is done for various values of $\phi$.
As the value of wave vector is small, calculated data points almost
show no angle $\phi$ dependence with the vertical scale used in Fig.4b.

\begin{figure}[ht]
\includegraphics[angle=90, width=8.truecm]{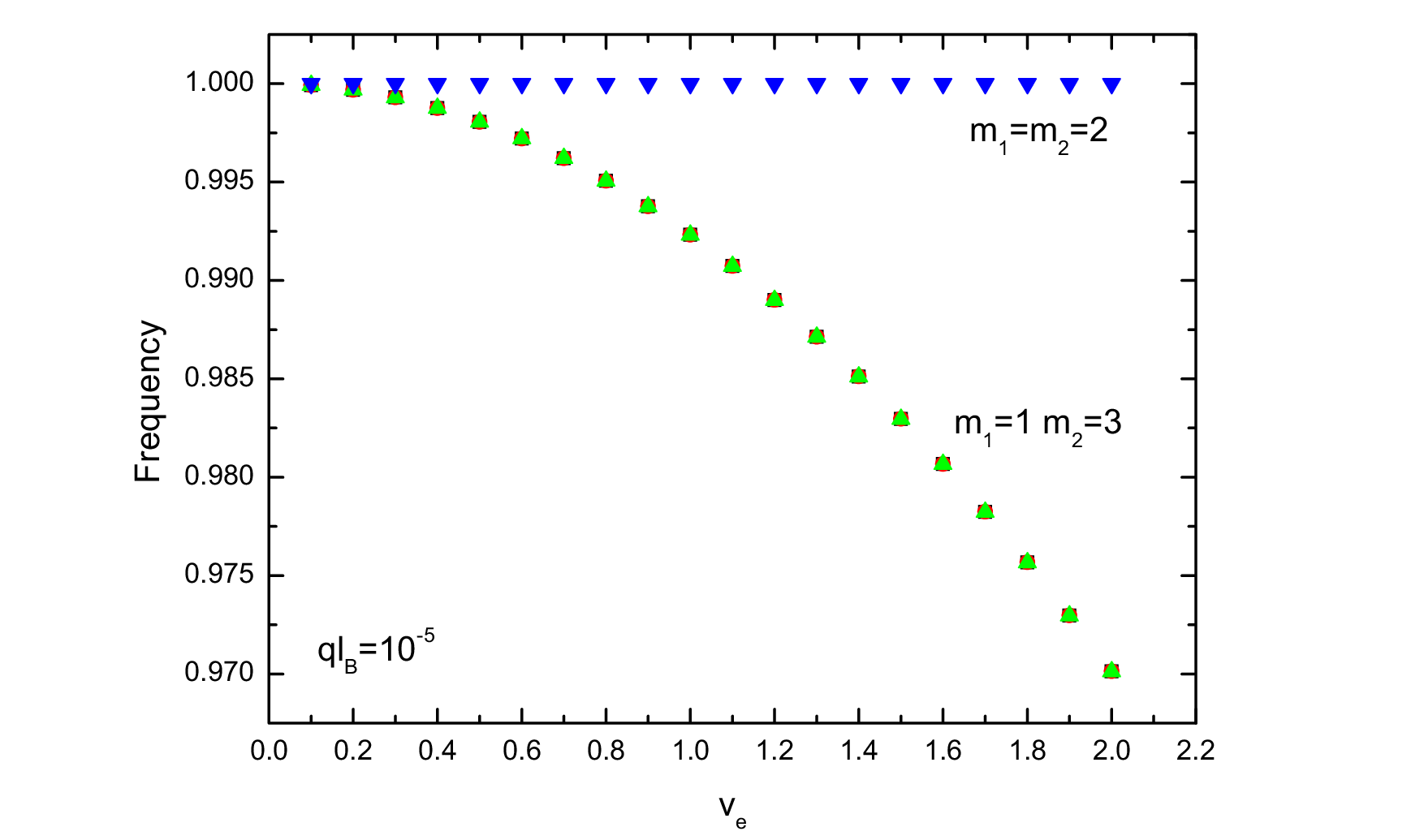}
\caption{
The excitation frequency versus the interaction strength $v_e$.
The filling factor is $1$. }
\end{figure}

In Fig.5, the excitation frequency is shown as a function of $v_e$, the
electron-electron interaction strength.  Parameters used in the
calculation are depicted in the figure.  The wave vector is fixed
at $ql_B=10^{-5}$.  The filling factor is $1$.

In the case of $m_1=m_2$, one observes that, the
scaled excitation frequency is independent of $v_e$.  In the case
of $m_1=1$ and $m_2=3$, the excitation frequencies are smaller than
$\omega_c$, and decrease as $v_e$ increases.  Various $\phi$ values
are used in the calculation for $m_1 \ne m_2$.  However, data points
are quite close to each other, almost overlapping with each other,
as the value of wave vector is very small.
It is quite clear that, in the case of $m_1 \ne m_2$, the $v_e$
dependence is not a linear one.

For a free 2D electron system, i.e., there is no electron-electron
interaction, the energy levels are degenerated Landau levels.  Two
adjacent Landau levels have the same distance $\hbar\omega_c$.
When the exchange effect is taken into account via the self-consistent
Hartree-Fock approximation, the gap between two adjacent Landau levels
(of the same spin)
becomes larger than $\hbar\omega_c$.
In the asymmetric 2D electron system studied here, the energy gap
induced by the exchange interaction reaches its maximal value
when $m_1=m_2$ \cite{xgwu}.
In the random phase approximation, using dressed Green's functions,
the excitation energy near the cyclotron resonance frequency will also
be larger than $\hbar\omega_c$.  The ladder diagram approximation would
reduce this excitation energy in the long wave length limit.
The calculated results shown in Fig.5 suggest that this reduction
or compensation is not large enough, when $m_1 \ne m_2$.

By examining the results presented above, one may conclude that, when the
wave vector approaches zero, in the case of $m_1=m_2$, the excitation
frequency near $\omega_c$ approaches $\omega_c$, but in the case
of $m_1 \ne m_2$, the excitation frequency approaches a value
smaller than $\omega_c$.  The Kohn's theorem \cite{kohn} requires that the
excitation frequency near $\omega_c$ should always approach $\omega_c$
in the long wave length limit.  Therefore, the approximation scheme used in
the current study fails, in the case of $m_1 \ne m_2$.
One should go beyond the approximation scheme adopted in the present study
to remedy the discrepancy.

\begin{figure}[ht]
\includegraphics[angle=0, width=8.truecm]{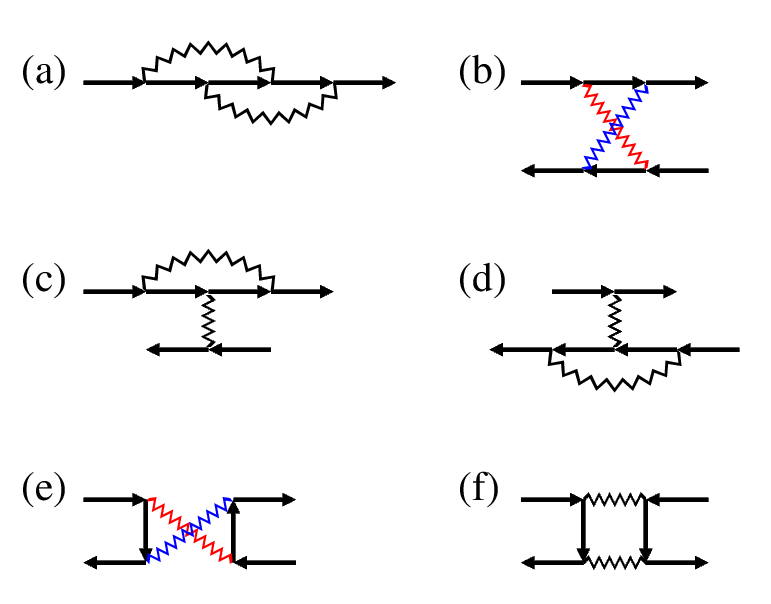}
\caption{
Some diagrams for the Green's function and the irreducible electron density
correlation function.}
\end{figure}

In the present study, the self-consistent Hartree-Fock approximation is
used in the calculation of Green's function.  This approximation is
employed for simplicity.  The irreducible polarization function is
calculated in the ladder diagram approximation.  The ladder diagrams are
also present in the self-consistent Hartree-Fock approximation, but the
Green's functions involved are different in two approximations.
In terms of unperturbed Green's functions, the self-consistent Hartree-Fock
approximation contains far more skeletons.
In the self-consistent Hartree-Fock approximation, the self-energy is
frequency independent, and this makes the calculation simpler.

The Hartree-Fock approximation is known to have some undesired effect.
In the zero magnetic field case, in a three-dimensional and 2D electron gas
system, the effective mass or self-energy shows a discontinuous or non-smooth behavior.
The problem can be circumvented by using the random phase approximation \cite{fetter, mahan}.
For a 2D electron gas in the presence of a perpendicular magnetic field,
the Hartree-Fock approximation is widely employed.  In this approximation,
the correlation effect is ignored.  The physics in the well-known fractional
quantum Hall effect is not captured by the Hartree-Fock approximation.
On the other hand, some magneto-optical experiments can be well described by simple theories
with no electron correlation effect being taken into account \cite{batke}.

\begin{figure}[ht]
\includegraphics[angle=90, width=8.truecm]{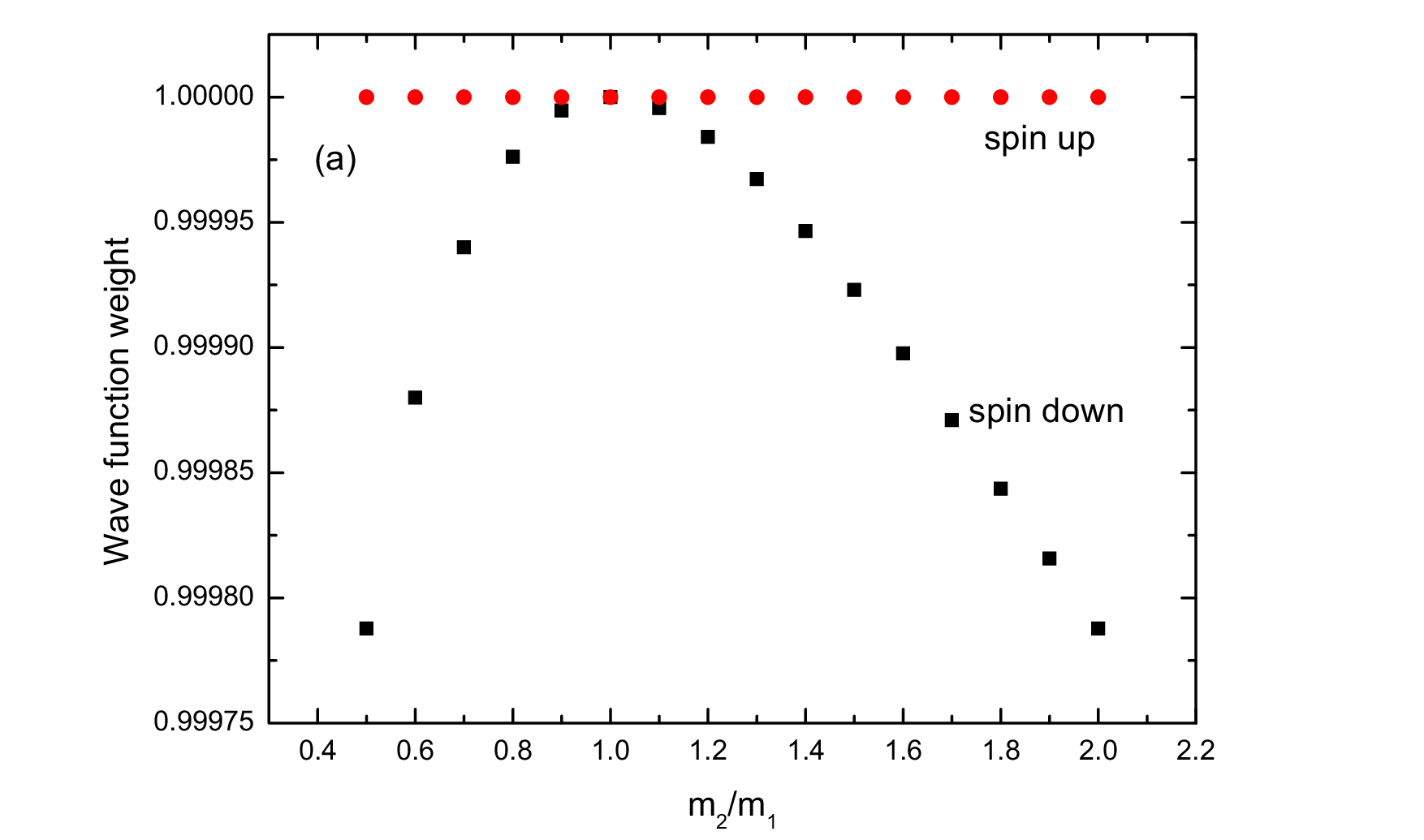}
\includegraphics[angle=90, width=8.truecm]{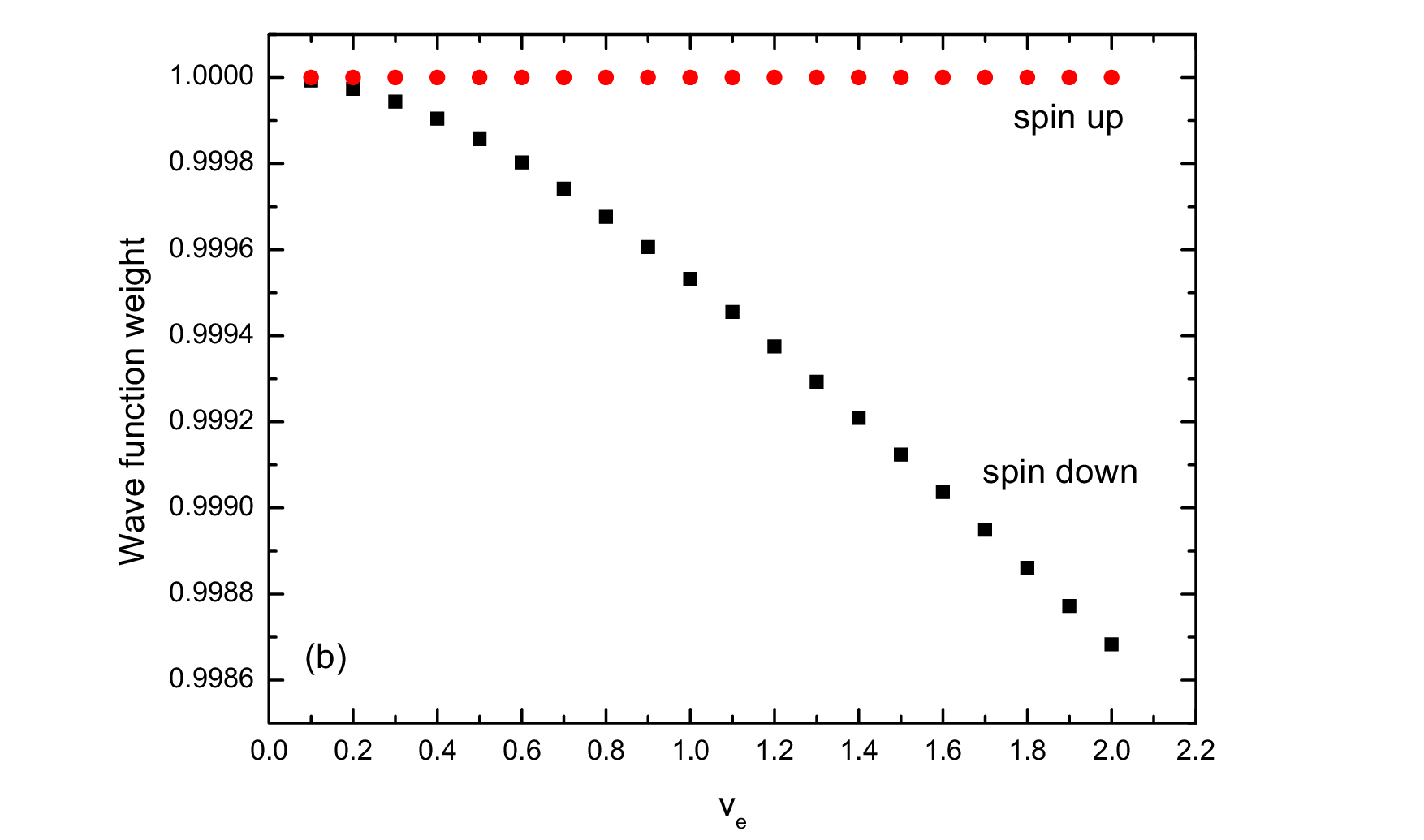}
\caption{
(a) The wave function weight versus $m_2/m_1$ with $v_e=1$.
(b) The wave function weight versus $v_e$ with $m_1=1$ and $m_2=3$.
Filling factor is $1$.
The wave function weight is defined as the lowest energy
level state projected onto the lowest single-particle Landau level.}
\end{figure}

The approximations used in this paper should be improved.
One possible route is to include more diagrams in the calculation of
self-energy and of irreducible electron density correlation function.
In Fig.6, some diagrams that contribute to the Green's function
and the irreducible electron density correlation function are shown.
We have assessed those terms carefully.

The lowest order self-energy vertex term shown in Fig.6a is examined, and it is
found that the self-energy will become frequency dependent.  The calculation
will involve high dimensional numerical integrals that must be done with very
high accuracy.  The required computational resource seems not reachable to us.

Including more diagrams, shown in Fig.6b to Fig.6f, in evaluating the
contribution to the irreducible density correlation will also lead to high
dimensional numerical integrals.  It is also found that, higher order
crossing diagram terms, i.e., like the one shown by Fig.6b with more interaction lines,
will lead to even higher dimensional numerical integrals.  The dimensionality
of the integrals will increase as the interaction order increases.
However, the task should be not impossible, we believe, if enough
computational resources were available.

One may keep the Green's function calculated in the
self-consistent Hartree-Fock approximation, and add some diagrams in
the evaluation of irreducible electron density correlation function.
The reason of doing so is given below.  The difference between calculated
excitation frequency and $\omega_c$ at small wave vectors, shown in
Figs.3-5, is actually quite small, much less than parameters $v_e$ or $m_2/m_1-1$.
In the self-consistent Hartree-Fock approximation, Landau levels are
mixed due to the asymmetry in our model system.
This Landau level mixing can be clearly seen by
checking the wave function weight, defined as the lowest energy level
state projected onto the lowest unperturbed Landau level.  If there is no mixing,
the weight will be $1$.  The mixing will reduce the weight.
In Fig.7a and Fig.7b, the wave function weight is shown as a function
of $m_2/m_1$ and $v_e$, respectively.  The filling factor is $1$.
Thus in this case, the spin-up state is not affected.  Only the spin-down
state is modified by the exchange effect of 2D interacting electrons.
One observes a clear similarity between Fig.4b and Fig.7a, and between
Fig.5 and Fig.7b.  This suggests that, due to Landau level mixing, the
ladder diagram contribution is slightly reduced, and leads to the
discrepancy shown in Figs.3-5.  By adding some diagrams to the ladder diagrams,
one may remove this discrepancy.

%
% summary
%
\section{Summary}

In summary, we have examined a GW-BSE approximation scheme applied to
an asymmetric 2D interacting electron system.
In this system, the well-known Kohn's theorem is still true.  We calculate
the Green's function in the self-consistent Hartree-Fock approximation, and
evaluate the electron density correlation function via solving a Bethe-Salpeter
equation in the ladder diagram approximation.  It is found that, the excitation
frequency near the cyclotron resonance frequency does not approach the cyclotron
resonance frequency at small wave vectors when two effective mass parameters
are different.  This is in contradiction with Kohn's theorem, and the
approximations scheme used should be improved.

%
% references
%


\begin{thebibliography}{}
\bibitem{note}
The present study focuses on small wave vectors.  In an earlier
paper \cite{xgwu}, the same model system was studied, and the excitation
energy at large wave vectors was evaluated and displayed.
However, the failure to satisfy the Kohn's theorem was not reported.
\end{thebibliography}
\end{document}